

\input harvmac

\noblackbox
\baselineskip 20pt plus 2pt minus 2pt

\overfullrule=0pt



\def\bs{\bigskip}
\def\no{\noindent}
\def\hb{\hfill\break}
\def\qq{\qquad}
\def\bl{\bigl}
\def\br{\bigr}

\def\IR{\relax{\rm I\kern-.18em R}}

\def\np {  Nucl. Phys. }


\def\a{\alpha}

\def\d{\delta}

\def\e{\epsilon}

\def\m{\mu}
\def\n{\nu}

\def\s{\sigma}

\def\IR{\relax{\rm I\kern-.18em R}}

\def \bd {\bar \del}

\def \ha {{1\over 2}}

\def \ov {\over}


\lref\BSthree{I. Bars and K. Sfetsos, Mod. Phys. Lett. {\bf A7} (1992) 1091.}

\lref\BShet{I. Bars and K. Sfetsos, Phys. Lett. {\bf 277B} (1992) 269.}

\lref\BSglo{I. Bars and K. Sfetsos, Phys. Rev. {\bf D46} (1992) 4495.}

\lref\BSexa{I. Bars and K. Sfetsos, Phys. Rev. {\bf D46} (1992) 4510.}

\lref\SFET{K. Sfetsos, Nucl. Phys. {\bf B389} (1993) 424.}

\lref\BSslsu{I. Bars and K. Sfetsos, Phys. Lett. {\bf 301B} (1993) 183.}

\lref\BSeaction{I. Bars and K. Sfetsos, Phys. Rev. {\bf D48} (1993) 844.}

\lref\BN{ I. Bars and D. Nemeschansky, Nucl. Phys. {\bf B348} (1991) 89.}

\lref\WIT{E. Witten, Phys. Rev. {\bf D44} (1991) 314.}

 \lref\IBhet{ I. Bars, Nucl. Phys. {\bf B334} (1990) 125. }

 \lref\IBCS{ I. Bars, {\it Curved Space-time Strings and Black Holes},
in Proc.
 {\it XX$^{th}$ Int. Conf. on Diff. Geometrical Methods in Physics}, eds. S.
 Catto and A. Rocha, Vol. 2, p. 695, (World Scientific, 1992).}

 \lref\CRE{M. Crescimanno, Mod. Phys. Lett. {\bf A7} (1992) 489.}

\lref\clapara{K. Bardakci, M. Crescimanno and E. Rabinovici,
Nucl. Phys. {\bf B344} (1990) 344.}

\lref\MSW{G. Mandal, A. Sengupta and S. Wadia,
Mod. Phys. Lett. {\bf A6} (1991) 1685.}

 \lref\HOHO{J. B. Horne and G. T. Horowitz, Nucl. Phys. {\bf B368} (1992) 444.}

 \lref\FRA{E. S. Fradkin and V. Ya. Linetsky, Phys. Lett. {\bf 277B}
          (1992) 73.}

 \lref\ISH{N. Ishibashi, M. Li, and A. R. Steif,
         Phys. Rev. Lett. {\bf 67} (1991) 3336.}

 \lref\HOR{P. Horava, Phys. Lett. {\bf 278B} (1992) 101.}

 \lref\RAI{E. Raiten, ``Perturbations of a Stringy Black Hole'',
         Fermilab-Pub 91-338-T.}

 \lref\GER{D. Gershon, ``Exact Solutions of Four-Dimensional Black Holes in
         String Theory'', TAUP-1937-91.}

 \lref \GIN {P. Ginsparg and F. Quevedo,  Nucl. Phys. {\bf B385} (1992) 527. }

 \lref\HOHOS{ J. H. Horne, G. T. Horowitz and A. R. Steif, Phys. Rev. Lett.
 {\bf 68} (1991) 568.}

 \lref\groups{
 M. Crescimanno. Mod. Phys. Lett. {\bf A7} (1992) 489. \hb
 J. B. Horne and G.T. Horowitz, Nucl. Phys. {\bf B368} (1992) 444. \hb
 E. S. Fradkin and V. Ya. Linetsky, Phys. Lett. {\bf 277B} (1992) 73. \hb
 P. Horava, Phys. Lett. {\bf 278B} (1992) 101.\hb
 E. Raiten, ``Perturbations of a Stringy Black Hole'',
         Fermilab-Pub 91-338-T.\hb
 D. Gershon, ``Exact Solutions of Four-Dimensional Black Holes in
         String Theory'', TAUP-1937-91.}

\lref\NAWIT{C. Nappi and E. Witten, Phys. Lett. {\bf 293B} (1992) 309.}

\lref\FRATSE{E. S. Fradkin and A. A. Tseytlin,
Phys. Lett. {\bf 158B} (1985) 316.}

\lref\CALLAN{ C. G. Callan, D. Friedan, E. J. Martinec and M. Perry,
Nucl. Phys. {\bf B262} (1985) 593.}

\lref\DB{L. Dixon, J. Lykken and M. Peskin, Nucl. Phys.
{\bf B325} (1989) 325.}

\lref\IB{I. Bars, Nucl. Phys. {\bf B334} (1990) 125.}

\lref\BUSCHER{T. Buscher, Phys. Lett. {\bf 201B} (1988) 466.}

\lref\Dual{M. Rocek and E. Verlinde, Nucl. Phys. {\bf B373} (1992) 630.\hb
A. Giveon and M. Rocek, Nucl. Phys. {\bf B380} (1992) 128. }

\lref\nadual{X. C. de la Ossa and F. Quevedo, ``Duality Symmetries from
Non-abelian Isometries in String Theory'', NEIP-92-004.}

\lref\SENrev{A. Sen, ``Black Holes and Solitons in String Theory'',
TIFR-TH-92-57.}

\lref\TSEd{A. A. Tseytlin, Mod. Phys. Lett. {\bf A6} (1991) 1721.}

\lref\TSESC{A. S. Schwarz and A. A. Tseytlin, ``Dilaton shift under duality
and torsion of elliptic complex'', IMPERIAL/TP/92-93/01. }

\lref\Dualone{K. Meissner and G. Veneziano,
Phys. Lett. {\bf B267} (1991) 33. \hb
K. Meissner and G. Veneziano, Mod. Phys. Lett. {\bf A6} (1991) 3397. \hb
M. Gasperini and G. Veneziano, Phys. Lett. {\bf 277B} (1992) 256. \hb
M. Gasperini, J. Maharana and G. Veneziano, Phys. Lett. {\bf 296B} (1992) 51.}

\lref\Dualtwo{A. Sen,
Phys. Lett. {\bf B271} (1991) 295;\ ibid. {\bf B274} (1992) 34. \hb
A. Sen, Phys. Rev. Lett. {\bf 69} (1992) 1006. \hb
S. Hassan and A. Sen, Nucl. Phys. {\bf B375} (1992) 103. \hb
J. Maharana and J. H. Schwarz, Nucl. Phys. {\bf B390} (1993) 3.}

\lref\KIRd{E. Kiritsis, ``Exact Duality Symmetries in CFT and String Theory'',
LPTENS-92-29; CERN-TH-6797-93.}

\lref\GIPA{A. Giveon and A. Pasquinucci, ``On cosmological string backgrounds
with toroidal isometries'', IASSNS-HEP-92/55, August 1992.}

\lref\KASU{Y. Kazama and H. Suzuki, Nucl. Phys. {\bf B234} (1989) 232. \hb
Y. Kazama and H. Suzuki Phys. Lett. {\bf 216B} (1989) 112.}

\lref\WITanom{E. Witten, Comm. Math. Phys. {\bf 144} (1992) 189.}

\lref\WITnm{E. Witten, Nucl. Phys. {\bf B371} (1992) 191.}

\lref\IBhetero{I. Bars, Phys. Lett. {\bf 293B} (1992) 315.}

\lref\IBerice{I. Bars, {\it Superstrings on Curved Space-times}, Lecture
delivered at the Int. workshop on {\it String Quantum Gravity and Physics
at the Planck Scale}, Erice, Italy, June 1992.}

\lref\DVV{R. Dijkgraaf, E. Verlinde and H. Verlinde, Nucl. Phys. {\bf B371}
(1992) 269.}

\lref\TSEY{A. A. Tseytlin, Phys. Lett. {\bf 268B} (1991) 175.}

\lref\JJP{I. Jack, D. R. T. Jones and J. Panvel,
          Nucl. Phys. {\bf B393} (1993) 95.}

\lref\BST { I. Bars, K. Sfetsos and A. A. Tseytlin, unpublished. }

\lref\TSEYT{ A. A. Tseytlin, ``Effective Action in Gauged WZW Models
and Exact String Solutions", Imperial/TP/92-93/10.}

\lref\TSEYTt{A. A. Tseytlin, ``Conformal Sigma Models corresponding to Gauged
WZW Models'', CERN-TH.6804/93.}

 \lref\SHIF { M. A. Shifman, Nucl. Phys. {\bf B352} (1991) 87.}
\lref\SHIFM { H. Leutwyler and M. A. Shifman, Int. J. Mod. Phys. {\bf
A7} (1992) 795. }

\lref\POLWIG { A. M. Polyakov and P. B. Wiegman, Phys.
Lett. {\bf 141B} (1984) 223.  }

\lref\BCR{K. Bardakci, M. Crescimanno
and E. Rabinovici, Nucl. Phys. {\bf B344} (1990) 344. }

\lref\Wwzw{E. Witten, Commun. Math. Phys. {\bf 92} (1984) 455.}

\lref\GKO{P. Goddard, A. Kent and D. Olive, Phys. Lett. {\bf 152B} (1985) 88.}

\lref\Toda{A. N. Leznov and M. V. Saveliev, Lett. Math. Phys. {\bf 3} (1979)
489. \hb A. N. Leznov and M. V. Saveliev, Comm. Math. Phys. {\bf 74}
(1980) 111.}

\lref\GToda{J. Balog, L. Feh\'er, L. O'Raifeartaigh, P. Forg\'acs and A. Wipf,
Ann. Phys. (New York) {\bf 203} (1990) 76.; Phys. Lett. {\bf 244B}
(1990) 435.}

\lref\GWZW{ E. Witten, \np {\bf B223} (1983) 422. \hb
K. Bardakci, E. Rabinovici and B. S\"aring, Nucl. Phys. {\bf B299}
(1988) 157. \hb K. Gawedzki and A. Kupiainen, Phys. Lett. {\bf 215B}
(1988) 119. \hb K. Gawedzki and A. Kupiainen, Nucl. Phys. {\bf B320}
(1989) 625. }

\lref\SCH{ D. Karabali, Q-Han Park, H. J. Schnitzer and Z. Yang,
                   Phys. Lett. {\bf B216} (1989) 307. \hb D. Karabali
and H. J. Schnitzer, Nucl. Phys. {\bf B329} (1990) 649. }

 \lref\KIR{E. Kiritsis, Mod. Phys. Lett. {\bf A6} (1991) 2871. }

\lref\BIR{N. D. Birrell and P. C. W. Davies,
{\it Quantum Fields in Curved Space}, Cambridge University Press.}

\lref\WYB{B. G. Wybourn, {\it Classical Groups for Physicists }
(John Wiley \& sons, 1974).}

\lref\SANTA{R. Guven, Phys. Lett. {\bf 191B} (1987) 275.\hb
D. Amati and C. Klimcik, Phys. Lett. {\bf 219B} (1989) 443.\hb
G. T. Horowitz and A. R. Steif, Phys. Rev. Lett. {\bf 64} (1990) 260.\hb
A.A. Tseytlin, Nucl. Phys. {\bf B390} (1993) 153.; Phys. Rev. {\bf D47}
(1993) 3421.}

\lref\SANT{J. H. Horne, G. T. Horowitz and A. R. Steif,
Phys. Rev. Lett. {\bf 68} (1991) 568.}

\lref\PRE{J. Prescill, P. Schwarz, A. Shapere, S. Trivedi and F. Wilczek,
Mod. Phys Lett. {\bf A6} (1991) 2353.\hb
C. Holzhey and F. Wilczek, Nucl. Phys. {\bf B380} (1992) 447.}

\lref\HAWK{J. B. Hartle and S. W. Hawking Phys. Rev. {\bf D13} (1976) 2188.\hb
S. W. Hawking, Phys. Rev. {\bf D18} (1978) 1747.}

\lref\HAWKI{S. W. Hawking, Comm. Math. Phys. {\bf 43} (1975) 199.}

\lref\HAWKII{S. W. Hawking, Phys. Rev. {\bf D14} (1976) 2460.}

\lref\euclidean{S. Elitzur, A. Forge and E. Rabinovici,
Nucl. Phys. {\bf B359} (1991) 581. }

\lref\ITZ{C. Itzykson and J. Zuber, {\it Quantum Field Theory},
McGraw Hill (1980). }

\lref\kacrev{P. Goddard and D. Olive, Journal of Mod. Phys. {\bf A} Vol. 1,
No. 2 (1986) 303.}

\lref\BBS{F. A. Bais, P. Bouwknegt and M. Surridge, Nucl. Phys. {\bf B304}
(1988) 348.}

\lref\nonl{A. Polyakov, {\it Fields, Strings and Critical Phenomena}, Proc. of
Les Houses 1988, eds. E. Brezin and J. Zinn-Justin North-Holland, 1990.\hb
Al. B. Zamolodchikov, preprint ITEP 87-89. \hb
K. Schoutens, A. Sevrin and P. van Nieuwenhuizen, Proc. of the Stony Brook
Conference {\it Strings and Symmetries 1991}, World Scientific,
Singapore, 1992. \hb
J. de Boer and J. Goeree, ``The Effective Action of $W_3$ Gravity to all
\hb orders'', THU-92/33.}

\lref\HOrev{G. T. Horowitz, {\it The Dark Side of String Theory:
Black Holes and Black Strings}, Proc. of the 1992 Trieste Spring School on
String Theory and Quantum Gravity.}

\lref\HSrev{J. Harvey and A. Strominger, {\it Quantum Aspects of Black
Holes}, Proc. of the 1992 Trieste Spring School on
String Theory and Quantum Gravity.}

\lref\GM{G. Gibbons, Nucl. Phys. {\bf B207} (1982) 337.\hb
G. Gibbons and K. Maeda, Nucl. Phys. {\bf B298} (1988) 741.}

\lref\GID{S. B. Giddings, Phys. Rev. {\bf D46} (1992) 1347.}

\lref\PRErev{J. Preskill, {\it Do Black Holes Destroy Information?},
Proc. of the International Symposium on Black Holes, Membranes, Wormholes,
and Superstrings, The Woodlands, Texas, 16-18 January, 1992.}

\lref\tye{S-W. Chung and S. H. H. Tye, Phys. Rev. {\bf D47} (1993) 4546.}

\lref\eguchi{T. Eguchi, Mod. Phys. Lett. {\bf A7} (1992) 85.}

\lref\HSBW{P. S. Howe and G. Sierra, Phys. Lett. {\bf 144B} (1984) 451.\hb
J. Bagger and E. Witten, Nucl. Phys. {\bf B222} (1983) 1.}

\lref\GSW{M. B. Green, J. H. Schwarz and E. Witten, {\it Superstring Theory},
Cambridge Univ. Press, Vols. 1 and 2, London and New York (1987).}

\lref\KAKU{M. Kaku, {\it Introduction to Superstrings}, Springer-Verlag, Berlin
and New York (1991).}

\lref\LSW{W. Lerche, A. N. Schellekens and N. P. Warner, {\it Lattices and
Strings }, Physics Reports {\bf 177}, Nos. 1 \& 2 (1989) 1, North-Holland,
Amsterdam.}

\lref\confrev{P. Ginsparg and J. L. Gardy in {\it Fields, Strings, and
Critical Phenomena}, 1988 Les Houches School, E. Brezin and J. Zinn-Justin,
eds, Elsevier Science Publ., Amsterdam (1989). \hb
J. Bagger, {\it Basic Conformal Field Theory},
Lectures given at 1988 Banff Summer Inst. on Particle and Fields,
Banff, Canada, Aug. 14-27, 1988, HUTP-89/A006, January 1989. }

\lref\CHAN{S. Chandrasekhar, {\it The Mathematical Theory of Black Holes},
Oxford University Press, 1983.}

\lref\KOULU{C. Kounnas and D. L\"ust, Phys. Lett. {\bf 289B} (1992) 56.}

\lref\PERRY{M. J. Perry and E. Teo, ``Non-singularity of the Exact two
Dimensional String Black Hole'', DAMTP-R-93-1. \hb
P. Yi, ``Nonsingular 2d Black Holes and Classical String Backgrounds'',
CALT-68-1852. }

\lref\GiKi{A. Giveon and E. Kiritsis, ``Axial Vector Duality Symmetry and
Topology Change in String Theory'', CERN-TH-6816-93.}

\lref\kar{S. K. Kar and A. Kumar, Phys. Lett. {\bf 291B} (1992) 246.}

\lref\NW{C. Nappi and E. Witten, ``A WZW model based on a non-semi-simple
group'',
IASSNS-HEP-93/61, hepth/9310112.}

\lref\HK{M. B. Halpern and E. Kiritsis,
Mod. Phys. Lett. {\bf A4} (1989) 1373.\hb
A.Yu. Morozov, A.M. Perelomov, A.A. Rosly, M.A. Shifman and A.V. Turbiner,
Int. J. Mod. Phys. {\bf A5} (1990) 803.}

\lref\KK{E. Kiritsis and C. Kounnas, ``String propagation in Gravitational Wave
Backgrounds'',
CERN-TH.7059/93, hepth/9310202.}
\lref\KST{K. Sfetsos and A.A. Tseytlin, ``Antisymmetric tensor coupling and
conformal
invariance in sigma models corresponding to gauged WZNW theories'',
CERN-TH.6969/93,
THU-93/25, hepth/9310159.}

\lref\KSTh{K. Sfetsos and A.A. Tseytlin, ``Chiral gauged WZNW models and
heterotic string
backgrounds'', CERN-TH.6962/93, USC-93/HEP-S2, hepth/9308018, to appear in
Nucl. Phys.
{\bf B}.}

\lref\Etc{K. Sfetsos, ``Gauging a non-semi-simple WZW model'',
THU-93/30, hepth/9311010.}

\lref\KT{C. Klimcik and A.A. Tseytlin, ``Duality invariant class of
exact string backgrounds'', CERN TH.7069, hepth/911012.}

\lref\ORS{ D. I. Olive, E. Rabinovici and A. Schwimmer, ``A class of String
Backgrounds as a Semiclassical Limit of WZW Models'', SWA/93-94/15,
hepth/93011081.}


\hfill {THU-93/31}
\vskip -.3 true cm
\rightline{November 1993}
\vskip -.3 true cm
\rightline {hep-th/9311093 }

\bs\bs\bs

\centerline  {\bf EXACT STRING BACKGROUNDS FROM WZW MODELS   }
\centerline  { \bf BASED ON NON-SEMI-SIMPLE GROUPS }

\vskip 1.00 true cm

\centerline  {  {\bf Konstadinos Sfetsos}{\footnote{$^*$}
 {e-mail address: sfetsos@ruunts.fys.ruu.nl }}
   }

\bigskip

\centerline {Institute for Theoretical Physics }
\centerline {Utrecht University}
\centerline {Princetonplein 5, TA 3508}
\centerline{ The Netherlands }


\vskip 1.50 true cm

\centerline{ABSTRACT}

\vskip .3 true cm

We formulate WZW models based on a centrally extended version of the
Euclidean group in $d$-dimensions. We obtain string backgrounds corresponding
to conformal $\s$-models in $D=d^2$ space-time dimensions with exact
central charge $c=d^2$ and $d(d-1)/2$ null Killing vectors.
By identifying the corresponding conformal field theory we show that
the one loop results coincide with the exact ones up to a shifting
of a parameter.

\vfill\eject


\newsec{Introduction}

Recently some attention was given to the construction of string backgrounds
from WZW models based on non-semi-simple groups. These provide new classes
of exactly solvable models and they seem to correspond to plane wave-type
with null Killing vectors solutions to string theory \SANTA.
Writing the WZW action corresponding to a non-semi-simple
group is not completely
straightforward since the quadratic form one usually constructs from the
algebra structure constants, namely $\Omega_{AB}=f_{AC}{}^D f_{BD}{}^C$,
is not invertible. Nappi and Witten in \NW\ showed how to circumvent
this problem for the particular case of the Euclidean group in two dimensions
$E_2$ by considering instead $E_2^c$, i.e. a centrally extended version of it.
The corresponding 4-dimensional string background, with exact central charge
$c=4$, is of the plane wave-type
and is exact to all orders in perturbation theory \NW.
The representation theory (and a complete bosonization) for $E_2^c$ was worked
out \KK\ and string
backgrounds in $D=3$ were constructed by gauging various 1-dimensional
subgroups of $E_2^c$ \KK\Etc.
It has also been shown that the background of \NW\
corresponds to a larger class of 1-loop solutions to string theory with
a null Killing vector which are also exact \KT.

In this article we generalize the work of \NW\ to a larger class of WZW models
based on a central extension of the Euclidean group in d-dimensions $E_d$,
which we will accordingly denote by $E_d^c$.
The formulation as WZW models and the
underlying current algebra symmetry makes them exactly solvable and one can
in principle compute their spectrum using current algebra techniques.

This paper is organized as follows: In Section 2 we work out explicitly the
case of the WZW model based on $E_d^c$.
The corresponding $D=d^2$-dimensional conformal $\s$-model possesses
$d(d-1)/2$ null Killing vectors.
In Section 3 we identify the corresponding conformal field theory (CFT)
(with central charge $c=d^2$) and show that
the 1-loop results for the non-linear $\s$-model corresponding to the WZW
action of the previous section are in fact exact to all orders (up to a
shifting of a parameter). This fact
can be traced back to the existence of $d(d-1)/2$ null Killing vectors as it
will be discussed. For simplicity some of the explicit computations
of this section are given
only for the case of $E_3^c$ but the final conclusions are true for the general
case as well. Section 4 contains discussion and concluding remarks.


\newsec { WZW action based on $E_d^c$}

The Euclidean group in $d$-dimensions has $d(d+1)/2$ generators
$\{J_{ij}, P_i\}\ ,i=1,2\dots d$. The commutation rules of the
corresponding algebra are

\eqn\ed{ [J_{ij},J_{kl}]=i \d_{[i[k} J_{j]l]}\ ,\qq
[J_{ij},P_k]=i \d_{[ik}P_{j]}\ ,\qq [P_i,P_j]=0\ .}
Obviously \ed\ is not a semi-simple algebra because it contains the abelian
ideal $\{P_i\}$ and can be thought as the semi-direct sum of
the algebra for the $SO(d)$ group and the group of translations in
$d$-dimensions, i.e. $so(d)\oplus_s T_d$.
If one attempts to write down a WZW action using the quadratic form
corresponding to the Killing metric, i.e. $\Omega_{AB}=f_{AC}{}^D f_{BD}{}^C$,
one discovers that this is reduced to the WZW model for
the $d(d-1)/2$-dimensional group $SO(d)$.
One can centrally extend \ed\ to $E_d^c$ by adding
the set of generators $\{T_{ij}\}$. The additional set of commutation rules
which preserve the Jacobi identities are

\eqn\edc{[J_{ij},T_{kl}]=i \d_{[i[k} T_{j]l]}\ ,\qq [P_i,P_j]=i T_{ij} }
and where the central extension operators $T_{ij}$
commute with everything else.
The quadratic form necessary to write down the corresponding WZW
action should satisfy the following criteria
a) $\Omega_{AB}=\Omega_{BA}$, b) $f^D_{AB}\Omega_{CD} +
f^D_{AC}\Omega_{BD} = 0$ and c) it should be  non-degenerate,
i.e. the inverse matrix
$\Omega^{AB}$ obeying $\Omega^{AB}\Omega_{BC}=\delta^A_C$ should exist.
The first and the second properties ensure the existence of the quadratic and
the Wess-Zumino term in the WZW action and the third one gives a way to
properly lower and raise indices.
For $E_d^c$ the unique solution that satisfies all of the above criteria is

\eqn\edcqf{\Omega_{AB}=\bordermatrix{& P_j & J_{kl} & T_{kl}\cr
P_i & \d_{ij} & 0 & 0\cr
J_{ij} & 0 & k\ \d_{[ik} \d_{j]l} & \d_{[ik} \d_{j]l} \cr
T_{ij} & 0 & \d_{[ik} \d_{j]l} & 0\cr}\ .}
The group element $g$ contains $d^2$ parameters and can be parametrized
as follows (where summation over repeated indices is implied)

\eqn\param{g=e^{i a\cdot P} e^{i v\cdot T} h_u\ ,\qq a\cdot P=a_iP_i\ ,
\qq v\cdot T=\ha v_{ij} T_{ij}\ ,}
where $v_{ij}=-v_{ji}$ and $h_u$ is an $SO(d)$ group element parametrized
in terms of $d(d-1)/2$ parameters $u_{ij}$.
Using the formula
$$ de^H=\int_0^1 dx\ e^{x H} dH e^{(1-x) H}$$
one can compute the following

\eqn\lj{g\inv dg= i(da\cdot P' -\ha da_i a_j T'_{ij})+i dv\cdot T'
+\ha (dh_u h_u\inv)_{ij} J'_{ij}\ ,}
where the generators $T'_A=h_u\inv T_A h_u$ satisfy the same commutation
relations \ed\ as the $T_A$'s. Similarly we compute
\eqn\rj{dgg\inv=
i(da\cdot P +\ha da_i a_j T_{ij})+i e^{ia\cdot P} dv\cdot T e^{-i\a\cdot P}
+e^{i\a\cdot P} e^{i\a\cdot v} dh_u h_u\inv
e^{-i\a\cdot v} e^{-i\a\cdot P} \ .}
In section 3 we will need an explicit expression for $dgg\inv$. One way to
obtain such an expression is to note that $dgg\inv=-gdg\inv$ and then use
\lj\ with ($a_i\to -a_i$, $ v_i\to -v_i$, $h_u\to h_u\inv$) and
($P_i\to P_i'$, $T_{ij}\to T'_{ij}$) with the additional
contribution of the terms $a\cdot dP'$ and $v\cdot dT'$ (these terms contribute
when derivatives with respect to the parameters $u_{ij}$ of $h_u$ are taken).
The resulting expression is
\eqn\dggi{\eqalign{dgg\inv=&i(da\cdot P +\ha da_i a_j T_{ij}) + idv\cdot T
+\ha (dh_uh_u\inv)_{ij} J_{ij} + a_i (dh_u h_u\inv )_{ij} P_j \cr
&+ (dh_u h_u)_{ij}(\ha a_i a_k  + v_{ik}) T_{jk}\ .
\cr }}

\no
The WZW action is defined as (we omit an overall scale factor)

\eqn\wzw{\eqalign{S(g)&=S_2(g)+S_3(g)\cr
&=-{1\ov 2\pi}\int_\Sigma d^2z\ T(g\inv \del g g\inv \bd g \Omega)
+{1\ov 6\pi}\int_B \Tr (g\inv dg \wedge g\inv dg \wedge g\inv dg \Omega )\ ,
\cr}}
where $\Omega_{AB}=Tr(T_A T_B \Omega)$ and $\Sigma=\del B$.
Using \lj, \edcqf\ and the fact that $h_u\inv \Omega h_u=\Omega$
one easily computes that

\eqn\stwo{S_2(g)=S_2(h_u)+{1\ov 4\pi} \int_{\Sigma} d^2z\
[2 \del a_i \bd a_i + i(\del h_u h_u\inv)_{ij}( \bd a_i a_j - \bd v_{ij})
+i( \del a_i a_j - \del v_{ij}) (\bd h_u h_u\inv)_{ij} ]\ .}
The computation of $S_3(g)$ is more involved. We will give some of the steps

\eqn\sth{\eqalign{&S_3(g)=
S_3(h_u) - {1\ov 6\pi} \int_B \Tr(da\cdot P'\wedge da \cdot P'
\wedge (dh_u h_u\inv)' \Omega ) \cr
&-{1\ov 6\pi} \int_B \Tr[\bl(da\cdot P' \wedge (dh_u h_u\inv)'
+ (dh_u h_u\inv)' \wedge  da\cdot P'\br)\wedge da\cdot P' \Omega ]   \cr
&+{i\ov 12\pi}\int_B \Tr[\bl((dv_{ij} -da_i a_j )T'_{ij} \wedge (dh_u h_u\inv)'
+(dh_u h_u\inv)' \wedge (dv_{ij}-da_i a_j ) T'_{ij} \br) \wedge (dh_u h_u\inv)'
\Omega ]\cr
& + {i\ov 12\pi} \int_B \Tr[ (dh_u h_u\inv)' \wedge (dh_u h_u\inv)' \wedge
(dv_{ij}-da_i a_j) T'_{ij} \Omega ]\cr
&= S_3(h_u) -{1\ov 4\pi} \int_B i(dh_u h_u\inv)_{ij} \wedge da_i \wedge da_j +
(dh_u h_u\inv)_{ij} \wedge (dh_u h_u\inv)_{il}
\wedge (dv_{jl}-\ha da_{[j} a_{l]} )\cr
&= S_3(h_u) + {i\ov 4\pi} \int_B d[ (dh_u h_u\inv)_{ij}\wedge
(dv_{ij}-da_i a_j) ]\cr
&= S_3(h_u) + {i\ov 4\pi} \int_{\Sigma} d^2z\ [ (\del h_u h_u\inv)_{ij}
(\bd v_{ij} - \bd a_i a_j ) -(\del v_{ij} - \del a_i a_j )
(\bd h_u h_u\inv)_{ij} ] \ ,\cr } }
where we used the definition $(dh_u h_u\inv)'=\ha (dh_u h_u\inv)_{ij} J'_{ij}$.
Combining \stwo\ with \sth\ we obtain the final form for the action

\eqn\wzes{S(g)=S(h_u) + {1\ov 2\pi} \int_{\Sigma} d^2z\ [ \del a_i \bd a_i
+ i(\del a_i a_j - \del v_{ij} ) ( \bd h_u h_u\inv)_{ij}   ] \ ,}
where $S(h_u)$ is the WZW action for $SO(d)$ at level $k$.
{}From the action \wzes\ one can easily read off the corresponding metric
and antisymmetric tensor fields.
For the purposes of Section 3 let us note that for $E_3^c$  \wzes\
can we rewritten as (In this case any antisymmetric matrix $A_{ij}$ can be
parametrized in terms of a vector $A_i$, i.e. $A_{ij}=\e_{ijk} A_k$)

\eqn\wztt{S(g)=S(h_u) + {1\ov 2\pi} \int_{\Sigma} d^2z\ [ \del a_i \bd a_i
+ (2 \del v_k - \e_{ijk} \del a_i a_j ) R^k_{\m} \bd u^{\m} ]\ ,}
where $dh_u h_u\inv =i J_k R^k_{\m} du^{\m}\ ,i=1,2,3\ ,\m=1,2,3$.
Finally, let us note that if $J_{ij}\to i\e_{ij} J$, $T_{ij}\to -i \e_{ij} T$,
$u_{ij}\to - \e_{ij} u$, $a_i \to -i a_i\ ,i=1,2$ and $k=-b$ one obtains
for $E_2^c$ the results of \NW.

\newsec { CFT corresponding to the WZW model for $E_d^c$ }

In this section we identify the CFT corresponding to the WZW action for
$E_d^c$. The OPE for the current algebra according to \ed, \edcqf\ are
(we concentrate on the holomorphic part only)

\eqn\OPE{\eqalign{&J_{ij} J_{kl}\sim {i\d_{[i[k} J_{j]l]}\ov z-w}
+ {k\ \d_{[ik} \d_{j]l}\ov (z-w)^2} \cr
&J_{ij}P_k \sim {i \d_{[ik} P_{j]} \ov z-w} \ ,
\qq P_iP_j \sim {iT_{ij}\ov z-w}+{\d_{ij}\ov (z-w)^2} \cr
&J_{ij} T_{kl}\sim {i\d_{[i[k} T_{j]l]}\ov z-w}
+ { \d_{[ik} \d_{j]l}\ov (z-w)^2}\ ,\qq T_{ij}T_{kl}\sim 0\ .\cr } }
 The corresponding stress energy tensor

\eqn\vir{ T=\ha :(P_iP_i+ J_{ij} T_{ij} -\ha (2d -3 +k) T_{ij} T_{ij} ): }
satisfies the Virasoro algebra with central charge $c=d^2$
and corresponds to a solution of the Master equation
of \HK. With respect to \vir\ all currents are primary fields with
conformal dimension one.
In the rest of this section we will show that the metric
corresponding to the WZW action \wzes\ can also be obtained via the operator
method \DVV\BSexa. For simplicity of the presentation we will concentrate on
the case of $E_3^c$.
It is convenient to express the zero modes of the holomorphic
currents in \OPE\ as first order differential operators.
{}From \rj\ rewritten for $E_3^c$ we compute the following matrix
defined as $dgg\inv=i dX^M  E_M{}^A T_A$, where $X^M=\{a_i,v_i,u_{\m}\}$

\eqn\ema{E_M{}^A=\bordermatrix{& P_j & J_j & T_j \cr
a_i & \d_{ij}& 0& -\ha \e_{ijk} a_k \cr
u_{\m} & \e_{jkl} R^k_{\m} a_l & R^j_{\m}
& \ha (a_j a_k -a_l a_l \d_{jk} -2 \e_{jlk} v_l) R^k_{\m} \cr
v_i & 0 & 0 & \d_{ij} \ } \ .}
Its inverse is given by

\eqn\eam{E_A{}^M=\bordermatrix{& a_j  & u_{\m}& v_j \cr
P_i & \d_{ij} & 0 & \ha \e_{ijk} a_k \cr
J_i & \e_{ijk} a_k & R^ {\m}_i & \e_{ijk} v_k \cr
T_i & 0 & 0 & \d_{ij} \cr } \ .}
The first order differential operators defined as
$J_A=i E_A{}^M \partial_M$ satisfy
the commutation relations \ed, \edc. Their explicit expressions are

\eqn\oper{\eqalign{
& J_{J_i}=i\e_{ijk} (a_k \partial_{a_j} + v_k \partial_{v_j})
+i R^{\m}_i \partial_{u_{\m}}\cr
&J_{P_i}= i\partial_i +{i\ov 2} \e_{ijk} a_k \partial_{v_j}\ ,\qq
J_{T_i}=i \partial_{v_i}\ .\cr } }
The metric and the dilaton can be deduced by comparing \DVV\BSexa
$$H T= (L_0 +\bar L_0) T $$
\no
with
$$HT= -{1\ov \sqrt{G}\ e^{\Phi}}
\partial_{M} \sqrt{G}\ e^{\Phi}\ G^{MN} \partial_{N} T \ ,$$
\no
where $H$ is the Hamiltonian of the corresponding CFT, $L_0$ and $\bar L_0$ are
the zero modes of the holomorphic and antiholomorphic stress energy tensors
and $T$ denotes tachyonic states of the theory annihilated by the positive
modes of the holomorphic and antiholomorphic currents.
One can show that the physical condition for closed strings $(L_0-\bar L_0)T=0$
is obeyed and therefore one need only consider the action of $L_0$ on $T$.
In this way one obtains a constant dilaton and the inverse metric

\eqn\metin{G^{MN}=\bordermatrix{& a_j & u_{\n} & v_j\cr
a_i & \d_{ij} & 0 & -\ha \e_{ijk} a_k\cr
u_{\m} & 0 & 0 & R^{\m}_j \cr
v_i & \ha \e_{ijk} a_k & R^{\n}_j &  ({1\ov 4} a\cdot a -k -3)\d_{ij}
-{1\ov 4} a_i a_j }\ ,}
which upon inverting gives the metric corresponding to the action \wztt, but
with $k\to k+3$. Therefore our result \wztt\ is exact up to the forementioned
shifting of $k$. For the general case of $E_d^c$ the result for the metric
one obtains with the use of the operator method coincides
with the one
corresponding to the action \wzes\ up to a shifting $k\to k+2d-3$ in \wzes.

\newsec{ Discussion and concluding remarks}

The space-time corresponding to \wzes\ is a $d^2$-dimensional one with
$d(d-1)/2$ null Killing vectors equal in number to the independent
components of the matrix $(v_{ij})$. The facts that the only result of quantum
effects is to shift $k$ in \wzes\ and that the central charge equals the
number of independent fields ($c=d^2$) can be traced back to the existence of
$d(d-1)/2$ null Killing vectors. Their role in \wzes\ is that of
Langrange multipliers which `freeze' out flactuations of the $SO(d)$ currents
unless sources are introduced for these currents and this is the reason
why all renormalization effects (shifting of $k$) occur only in the part
of \wzes\ corresponding
to the WZW model for the group $SO(d)$. It is worth understanding this point
from the beta functions point of view along the lines of \KT.

The action \wzes\ has the following `obvious' global symmetries
$$ h_u\to h_u \Lambda\ ,\qq \{ h_u\to S h_u, v\to S v S^T, a_i\to S_{ij} a_j\}
\ ,\qq v\to v+N$$
where $\Lambda$, $S$ are constant group elements of $SO(d)$ and $N$ is a
constant antisymmetric matrix. The total number of constant parameters is
$3d(d-1)/2$.
It is very likely that one can obtain the background corresponding to \wzes\
by appropriate $O({3\ov 2} d(d-1),{3\ov 2} d(d-1))$ transformations
performed on the background
corresponding to a flat space-time with zero antisymmetric tensor in $D=d^2$
space-time dimensions (The fact that $c=d^2$ is very suggestive).\foot
{For the $E_2^c$ case this was shown in \KK\KT.
The same is true for all models one obtains by gauging 1-dimensional
subgroups of $E_2^c$ \Etc.} One can also use similar transformations to
generate new string backgrounds.

The signature of the space-time is
determined by the sign of the eigenvalues of the quadratic form \edcqf.
One easily finds that there are $d(d+1)/2$ positive and $d(d-1)/2$
negative ones.
One may consider gauged versions of the WZW model corresponding to $E_d^c$
as it was done for the case of $E_2^c$ in \KK\Etc.
For instance one could gauge the
subgroup corresponding to $E_{d-1}^c$ (for $d=2$ this was done in \Etc).
That will give a model in $D=2d-1$ space-time dimensions with
$(d-1)$ time-like coordinates.
Clearly only for $d=2$ one obtains a space-time with 1-time
coordinate in both the $E_d^c$ and the $E_d^c/E_{d-1}^c$ cases.

Finally, we believe that it is worth considering WZW models
based on other semi-simple groups (in particular those with
1-time coordinate)
since they correspond to solvable models with exact conformal invariance.
They seem promising candidates for exact CFT theories corresponding to
plane wave-type with null Killing vectors solutions to string theory.

\bigskip\bigskip

\centerline{ {\bf Note added} }

After we finished this paper we received ref. \ORS\ where the problem of
constructing WZW models based on non-semi-simple groups was also considered.

\listrefs

\end

The Euclidean group in three dimensions $E_3^c$ has 6 generators
$\{J_i,P_i\}\ ,i=1,2,3$. The commutation rules for the corresponding algebra
are

\eqn\edt{[J_i,J_j]=i \e_{ijk} J_k\ ,\qq [J_i,P_j]=i \e_{ijk} P_j\ ,
\qq [P_i,P_j]=0\ }
This algebra is not semi-simple (it contains the abelian ideal $\{P_i\}$
and it can be thought as the semi-direct sum of the Lie algebra $so(3)$ and
the group of translations in $d=3$, i.e. $so(3)\oplus_s T_3$.
One can centrally extend \edt\ to $E_3^c$ by adding
the set of generators $\{T_i\}$. The additional set of commutation rules
which preserve the Jacobi identities are

\eqn\addt{[J_i,T_j]=i \e_{ijk} T_k\ ,\qq [P_i,P_j]=i \e_{ijk} T_k \ ,}
and where the central extension generators $\{T_i\}$
commute with everything else.
The quadratic form necessary to write down the corresponding WZW
action should satisfy the following criteria
a) $\Omega_{AB}=\Omega_{BA}$, b) $f^D_{AB}\Omega_{CD} +
f^D_{AC}\Omega_{BD} = 0$ and c) is non-degenerate, i.e. the inverse matrix
$\Omega^{AB}$ obeying $\Omega^{AB}\Omega_{BC}=\delta^A_C$ exists.
The first and the second properties ensure the existence of the quadratic and
the Wess-Zumino term in the WZW action and the third one gives a way to
properly lower and raise indices.
For $E_3^c$ there is a unique solution that satisfies all of the above criteria

\eqn\ft{\bordermatrix{&P_j & J_j & T_j \cr
P_i & \d_{ij} & 0 & 0\cr
J_i & 0 & k/2\ \d_{ij} & \d_{ij} \cr
T_i & 0 & \d_{ij} & 0 \cr}\ .}